\documentclass[
aip,
amsmath,amssymb,
reprint,
]{revtex4-1}
\usepackage{fancyhdr}
\usepackage{calc}
\usepackage{amssymb}
\usepackage{setspace}
\usepackage{amsfonts}
\usepackage[innercaption]{sidecap}
\usepackage{graphicx}
\usepackage{graphicx,epstopdf}
\usepackage{xcolor, soul}
\usepackage{dcolumn}
\usepackage{bm}
\usepackage{mathrsfs} 
\usepackage{amsmath} 
\usepackage[colorlinks=true,linkcolor=blue,citecolor=blue]{hyperref}%
\usepackage{fontenc}
\usepackage{float}
\usepackage{amsthm}
\usepackage{subfigure}
\usepackage{color}
\usepackage{ragged2e}
\usepackage{enumerate}
\usepackage{hyperref}

\begin{document}
\title{Extreme events in a complex network: interplay between degree distribution and repulsive interaction}

\author{Arnob Ray}
\email{arnobray93@gmail.com}
\address{Physics and Applied Mathematics Unit, Indian Statistical Institute, Kolkata 700108, India}

\author{Timo Br\"{o}hl}
\address{Department of Epileptology, University Hospital Bonn, Venusberg Campus 1, 53127 Bonn, Germany}
\address{Helmholtz‑Institute for Radiation and Nuclear Physics, University of Bonn, Nussallee 14‑16, 53115 Bonn, Germany}

\author{Arindam Mishra}
\address{Department of Physics, National University of Singapore, Singapore 117551}

\author{Subrata Ghosh}
\address{Center for Computational Natural Sciences and Bioinformatics, International Institute of Information Technology, Gachibowli, Hyderabad 500032, India}

\author{Dibakar Ghosh}
\address{Physics and Applied Mathematics Unit, Indian Statistical Institute,  Kolkata 700108, India}

\author{Tomasz Kapitaniak}
\address{Division of Dynamics, Lodz University of Technology, 90-924 Lodz, Poland}

\author{Syamal K. Dana}
\address{National Institute of Technology, Durgapur 713209, India}

\author{Chittaranjan Hens}
\address{Center for Computational Natural Sciences and Bioinformatics, International Institute of Information Technology, Gachibowli, Hyderabad 500032, India}

\date{\today}

\begin{abstract} 
The role of topological heterogeneity in the origin of extreme events in a network is investigated here. The dynamics of the oscillators associated with the nodes are assumed to be identical and influenced by mean-field repulsive interactions. An interplay of topological heterogeneity and the repulsive interaction between the dynamical units of the network triggers extreme events in the nodes when each node succumbs to such events for discretely different ranges of repulsive coupling. A high degree node is vulnerable to weaker repulsive interactions, while a low degree node is susceptible to stronger interactions. As a result, the formation of extreme events changes position with increasing strength of repulsive interaction from high to low degree nodes. Extreme events at any node are identified with the appearance of occasional large-amplitude events (amplitude of the temporal dynamics) that are larger than a threshold height and rare in occurrence, which we confirm by estimating
the probability distribution of all events. Extreme events appear at any oscillator near the boundary of transition from rotation to libration at a critical value of the repulsive coupling strength. To explore the phenomenon, a paradigmatic second-order phase model is used to represent the dynamics of the oscillator associated with each node. We make an annealed network approximation to reduce our original model and thereby confirm the dual role of the repulsive interaction and the degree of a node in the origin of extreme events in any oscillator.
\end{abstract}
\maketitle
\begin{quotation}
Studies of extreme events in dynamical networks, brain networks, power-grid, and road networks are important for searching any early warning signals for the purpose of taking any preemptive measures for mitigating any disastrous effect on the network. Even while designing an engineering network, precautionary measures can be undertaken to identify the sources of instabilities that may lead to extreme conditions. Heterogeneity in the parameter distribution of the dynamical nodes of a network is one such source of instabilities that may lead to the formation of extreme events in all-to-all coupled networks. We raise a question here what happens in a complex network in the presence of any instabilities? Does the degree distribution of the nodes play any role, particularly when the oscillators associated with the nodes interact repulsively? The most important question is which nodes are most susceptible to the formation of extreme conditions. We address this question using a network with a degree of heterogeneity of nodes and a second-order phase model to represent the dynamics of the nodes.
\end{quotation}


\maketitle


\section{Introduction}\label{sec1}
 A significant progress has been witnessed, in the last decade, in the understanding of extreme events, especially in deterministic dynamical systems \cite{ghil2011extreme, sapsis2018new, farazmand2019extreme, mishra2020routes, chowdhury2021review}. Extreme events have been observed in single systems and ensemble of interacting systems. 
It appears as significantly large deviations from the long-time average behavior of an observable of a system. The large deviations are rare and short-lasting, but recurrent.  Many examples of extreme events can be cited from nature such as floods, droughts, earthquakes, tsunamis hurricanes, cyclones, regime shifts in ecosystems \cite{folke2004regime, scheffer2003catastrophic}, harmful algal blooms \cite{anderson2012progress} and many more as they appear with devastating impact on life, infrastructure and economy \cite{jusup2022social, helbing2015saving}. The share market crashes and power blackouts are other examples of extreme events in human-made and engineering systems.
 
\par To develop a better understanding of extreme events, besides the data-driven analysis, a more comprehensive study using simple dynamical models is a necessity that may help explain the origin of extreme events in nature, which are high dimensional and difficult to define \cite{ghil2011extreme, farazmand2019extreme, chowdhury2021review}. In fact, in many dynamical systems, infrequent and recurrent comparatively high or low amplitude events appear in their temporal dynamics that have qualitative similarities with occasional large events as recorded in many real-world phenomena. A striking example is the ocean rogue waves that make a devastating impact on ships and the seamen on the high seas. It  draws the attention of the researchers \cite{kharif2003physical, akhmediev2009extreme} in the 1990s to find an explanation. A trend of research follows to explain the phenomenon of rogue waves using the nonlinear Schr\"odinger equation \cite{akhmediev2009extreme} and then to reproduce the behavior in the laboratory using an optical fiber \cite{solli2007optical}. A few years later, the appearance of rogue waves was exhibited  \cite{bonatto2011deterministic} in a laser experiment whose dynamics is governed by a deterministic dynamical model. Since then, several numerical as well as experimental investigations on extreme events have been performed in a variety of dynamical systems, such as laser models \cite{zamora2013rogue, mercier2015numerical, ray2019intermittent}, electronic circuit \cite{cavalcante2013predictability, de2016local, kingston2017extreme}, climate model \cite{ray2020enso}, mechanical systems \cite{sudharsan2021emergence, meiyazhagan2021model}, neuronal models \cite{mishra2018dragon, saha2017extreme} and other systems \cite{kumarasamy2018extreme, kumarasamy2022emergence}. Two issues have mainly been focused, (i) the origin of extreme events, and (ii) prediction. 

\par The important nonlinear processes \cite{farazmand2019extreme, mishra2020routes, chowdhury2021review} that are involved in the origin of extreme events in single nonlinear dynamical systems, have mostly been unraveled by this time. The dynamical system community arrives at a  general consensus  that a source of instability  \cite{sapsis2018new, babaee2016variational} such as a saddle point, saddle orbit, unstable periodic orbit, or any form of singularity in the state space of a dynamical system, must be present in the phase space that is responsible for triggering extreme events. Whenever the trajectory of a system arrives at a distance close to any region of the instabilities, it deviates largely from its usual bounded region of the phase space, but returns to it after a short while. These occasional short-duration large deviations of the trajectory indicate the formation of  extreme events.  
A number of nonlinear processes that are involved in the origin of extreme events in dynamical systems, have already been reported such as interior crisis-induced intermittency \cite{zamora2013rogue, kingston2017extreme, bonatto2017extreme, ray2019intermittent, ray2020enso, thangavel2021extreme}, Pomeau–Manneville
intermittency \cite{mishra2020routes, kingston2017extreme,  kaviya2020influence}, quasiperiodic intermittency \cite{kingston2021instabilities}. Noise induced intermittency \cite{pisarchik2011rogue} is a very common source of instability that may originate extreme events. A specific system dependent source has also been identified such as the sliding bifurcation \cite{suresh2020parametric} that  triggers extreme events in the system, however, the list of mechanisms is not exhaustive.
\par
 Extreme events, in general, have its manifestation in intermittent large events that are characterized as chaotic in low dimensional systems, although they appear as more complex. Lately, extreme events are identified as hyperchaotic \cite{ mompo2021designing, kingston2022transition, leo2022transition} in higher dimensional systems. In coupled  systems,  
the sources of instabilities have also been explored \cite{cavalcante2013predictability,  mishra2018dragon, chowdhury2021extreme, moitra2019emergence, chowdhury2019synchronization, chowdhury2020distance}; some new routes are found, like on-off intermittency \cite{heagy1994characterization}. However, the fundamental rule of the presence of any sources of instability in phase space for the origin of extreme events remains always true. The main concern is how large are the events and whether the arrival of such complex temporal behavior is destructive to a system, natural or engineered. Thus the important task is to predict extreme events to mitigate any disastrous event, and it is always challenging and yet to be resolved since chaos or hyperchaos, it is always difficult to predict the future trajectory for a long time. In many cases, we use the dynamical behavior (like instability at the phase space) of the system for detecting an upcoming extreme event for prediction \cite{cavalcante2013predictability}. The data-driven methods (such as the machine learning technique \cite{tang2020introduction}) have also been attempted for the prediction of extreme events \cite{ray2021optimized, meiyazhagan2021model, meiyazhagan2022prediction, banerjeepredicting}. Along this line, {\it a  priori} attempts to include any appropriate control schemes during the design of engineering systems, which may help mitigate extreme events \cite{sudharsan2021constant, ray2019intermittent, sudharsan2021emergence}.

\par 
It is a pressing question of how to understand the origin of such devastating phenomena in dynamical networks. Some recent studies \cite{ansmann2013extreme, brohl2020identifying, ansmann2013extreme, werner2015transitions, rings2017important, ansmann2016self, brohl2020identifying, varshney2021traveling} have investigated the origin of extreme events in complex dynamical networks. Extreme events in ecological networks have also been reported \cite{moitra2019emergence, chaurasia2020advent}.
 Ansmann et al.\ \cite{ansmann2013extreme} have showed that extreme events appear in a network of excitable systems under attractive coupling when a critical number of nodes (each node represents an excitable neuron) of neurons fire coherently. This concept of local excitation of a fraction of dynamical units contributing to the formation of extreme events in a network has been confirmed \cite{brohl2020identifying} in a complex network of excitable systems. They have further extended the work to answer the question of how the fraction of coherently firing nodes are recruited \cite{brohl2019centrality}. In particular, the role of the edges has been recognized in selecting the particular fraction of nodes contributing to the formation of extreme events. Rings et al.\ \cite{rings2017important}, in another work, discussed the role of high degree and low degree nodes in the emergence of extreme events in  scale-free networks of excitable units and found that extreme events are initiated in the low degree nodes and then engulfed the whole network when hubs play a significant role. In this paper, we present a counterintuitive information about the initiation and spreading of extreme events in the oscillatory nodes of a complex network. Our results show that the hubs are intuitively the most vulnerable nodes and extreme events follow a spreading path by changing position from high degree nodes to the low degree nodes against an increasing repulsive interaction \cite{levnajic2011emergent, hens2013oscillation}.
 

\par In the past, we have explored \cite{ray2020extreme} globally coupled networks under repulsive interaction with heterogeneity in parameters of the governing dynamics of the oscillators associated with the nodes. An interplay of heterogeneity \cite{xu2021collective} of a system parameter and the repulsive interaction among the dynamical units leads to the triggering extreme events in a fraction of units of the network. We address a reverse question here what happens if we change the regular graph by inducing heterogeneity in the network topology and keeping the dynamical units identical? Does the degree supremacy of a node indulge any preference in the formation of extreme events in a complex network of identical oscillatory nodes? 
Indeed, we cannot avoid a situation when the extreme events preferentially appear in the higher degree nodes of the network for a range of weak repulsive coupling. The higher degree node is more susceptible to the formation of extreme events for a weaker repulsive coupling, while the smaller degree nodes remain dormant. Extreme events then start moving to the lower degree nodes when the repulsive interaction is increased, and surprisingly, then they stop appearing in the higher degree nodes at larger repulsive coupling. It is an established fact\cite{cohen2001breakdown,albert2000error} that an intentional attack, particularly, targeting the highest connectivity node in a scale-free network, can disrupt the network into small fragments. However, such networks are robust and resilient against a random attack, and thus hubs (large degree nodes) are the vulnerable nodes (even if the exponent of the degree distribution following power law is less than equal to $3$) in scale-free networks. In our case, a weak repulsive (negative) coupling  first destabilizes the original dynamics of the highest degree node. The dynamics of the hub (highest degree node) then remains confined, most of the time, into small amplitude oscillation, but occasionally transits to large amplitude oscillation (explained later). Once the repulsive coupling strength is increased, this occasional transition to large  amplitude oscillation is stopped in the hub and it becomes quiet. Then the next higher  degree node becomes vulnerable, showing a similar occasional transition to large amplitude oscillation, indicating a formation of extreme events, and so on. 
And thereby, the formation of extreme events moves from high degree nodes to low degree nodes with increasing the strength of the repulsive interaction.

\par We report our findings of numerical studies of the dynamics of the oscillators interacting in a landscape of a complex network. We use a paradigmatic second order phase model \cite{mishra2021neuron,  dana2006spiking} to represent the dynamics of the oscillator associated with each node. The phase model represents the dynamics of the superconducting Josephson junction \cite{dana2001chaotic}, a simple pendulum \cite{hongray2016dynamics},  and the discrete sine-Gordon equation \cite{rubinstein1970sine}. 
We describe the mathematical model of the network and the dynamics of identical oscillators in Sec.\ \ref{sec2}. We define the extreme events for our case study in Sec.\ \ref{def}. We describe the network structure and the dependency of  the formation of extreme events on repulsive coupling strength in Sec.\ \ref{sec3}. In this section, we present bifurcation diagrams to see the changes in dynamics against the repulsive interaction and the temporal patterns of extreme events. In Sec.\ \ref{sec4}, we use the annealed network approximation to reduce the large network model, and thereby explain how the interplay of degree of a node and the repulsive interaction leads to the formation of extreme events. Section\ \ref{sec5} deals with the statistical properties of extreme events and finally, we  make a conclusion in Sec.\ \ref{sec6}.
\section{Model description}\label{sec2}
We consider a model of coupled oscillators where underlying interaction topology is complex network with a heterogeneous degree distribution of the nodes.
The dynamics of the $i$-th node is represented by a second order phase model \cite{mishra2021neuron, strogatz2018nonlinear} with an external periodic forcing and influenced by a mean-field repulsive interaction as described below,
\begin{equation}
\begin{array}{l}\label{eq.1}	
\dot{\phi_{i}}  = y_{i},\\
\dot{y}_{i} = I - \sin \phi_{i} - \alpha y_{i} + I_f\sin (\Omega_f t) + \dfrac{K}{n}\sum_{j=1}^{n} A_{ij} {y_j},\\
~~~~~~~~~~~~~~~~~~~~~~~~~~~~~~~~~~~~~~~~~~~~~~~~~~~i=1,2,\cdots,n,
\end{array}
\end{equation}

 where $\phi_i$ is a phase variable,  $y_i$ is the phase velocity
and $\alpha$ 
is the damping parameter. $I$ denotes a constant quantity analogous to  constant bias current in the Josephson junction model \cite{mishra2021neuron, dana2006spiking, dana2001chaotic} or the torque in a pendulum motion \cite{hongray2016dynamics}. The amplitude and frequency of the forcing signal are represented by $I_{f}$, and $\Omega_{f}$, respectively, that adds complexity in the dynamics of the  oscillators. Here the adjacency matrix ${[A_{ij}]}_{n\times n}$ of a simple undirected network (consisting $n$ number of nodes) is defined as\\
\[
{A}_{ij}= 
\begin{cases}
1, & \text{if}~~i\text{-th node is connected with} j\text{-th node}, \\
0,              & \text{otherwise}.
\end{cases}
\]\\
$K$ defines the strength of repulsive mean-field interaction between the $i^{th}$ oscillator and the oscillators of the neighboring nodes in the  network. For this study, we always choose $K<0$ so that the mean-field interaction is made repulsive. The individual phase model shows a variety of dynamics consisting of mainly two kinds of motion \cite{mishra2021neuron, strogatz2018nonlinear}, (a) libration, and (b) rotation. In a $\phi$-$y$ cylindrical phase space, the trajectory may librate in a small amplitude motion like the back and forth oscillation of a pendulum  while in rotation, it traces a complete cycle around the cylindrical plane like an inverted pendulum. The rotational motion appears as large amplitude spikes in the time evolution of the $y$-variable.  This dynamical variation has striking similarities with spiking neurons \cite{dana2006spiking, mishra2021neuron} and it makes this particular model more encouraging to use in our proposed network. 

\par For our numerical investigations, we always take a network of size $n=200$ and fix the parameters of the phase  model at $I=1.2$, $\alpha=1.5$. The parameters of the external periodic signal are $I_{f} = 0.26$ and $\Omega_{f} = 0.4$. $K$ is chosen as the bifurcation parameter to explore the variation of dynamics and occurrence of extreme events. In uncoupled state, all the oscillators exhibit rotation \cite{ray2020extreme,  mishra2017coherent, hens2015bursting} for our choice of parameters. The initial values of $\phi_i$ and $y_i$ are chosen from a uniform distribution in such a way that $\phi_i=0.001 i$, and $y_i=0.0025 i$ where $i=1, 2,\dots,200$.
We generate the heterogeneous network using the Barab{\'a}si-Albert (BA) algorithm \cite{barabasi_science1999}.  We assume that each new node has $4$ new edges and we start the preferential attachment process from the initial network of size $4$. 
\begin{figure}[h]
	\centerline{\includegraphics[scale=0.64]{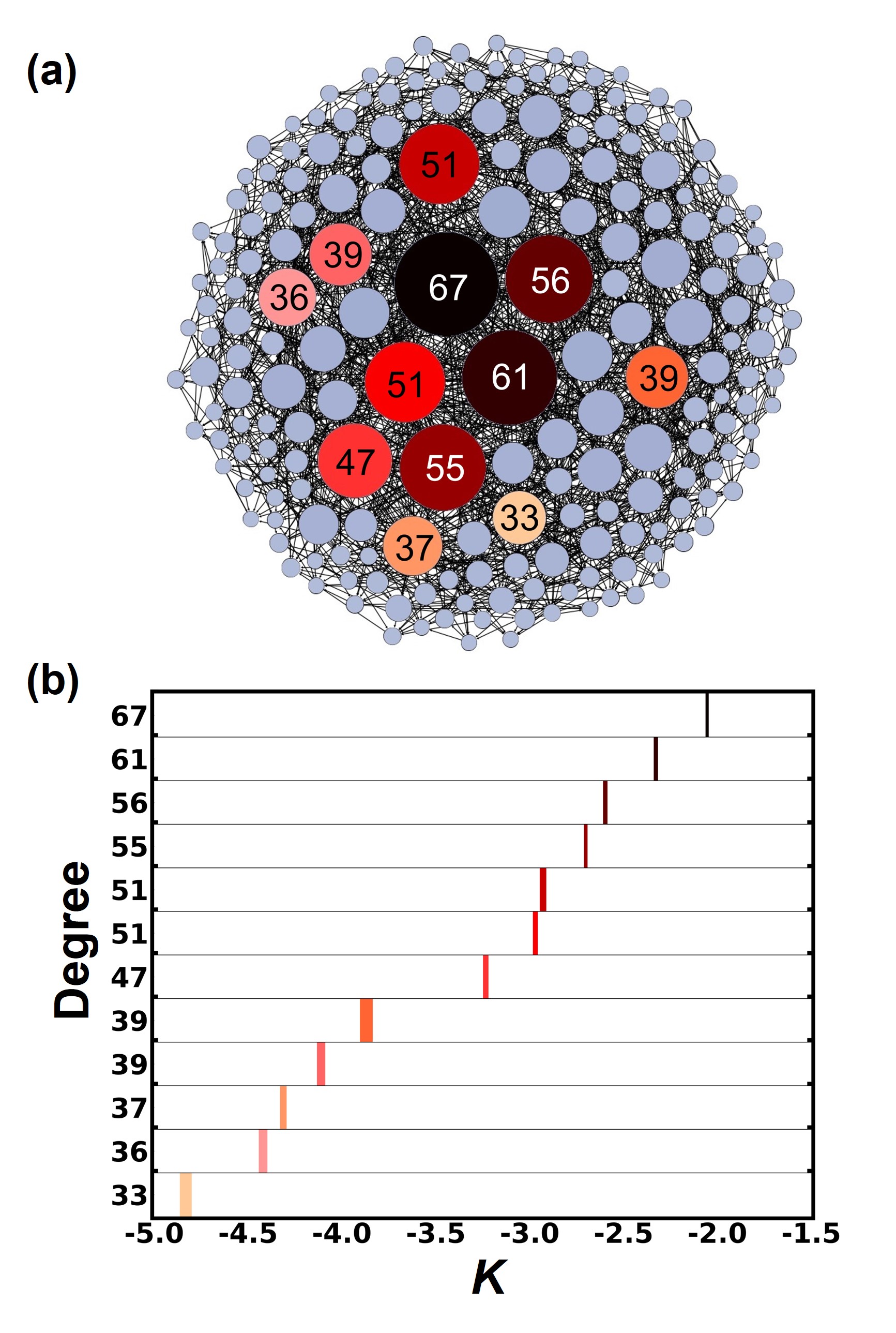}}
	\caption{Network structure and formation of extreme events. (a) We present a complex network that is chosen as the underlying network of your work, where a few nodes with high degrees are shown in circles of different colors. The larger the degree of a node is represented by larger radius. The numbers within circles indicate the degree of the nodes with different colors. (b) The ranges of $K$ where extreme events occur at the selective nodes are indicated by vertical color bars for different high degree nodes with variation of repulsive coupling strength when transition occurs between rotation to libration. The degree of the nodes is arranged in descending order along the ordinate from the top and the repulsive coupling strength is plotted along the abscissa. The width of the vertical color bars represents the range of coupling strength for which the extreme events are appearing at the particular node. The range of extreme events move from right to the left from the high to low degree nodes with increasing the magnitude of the repulsive strength $(|K|)$. Parameter values: $I=1.2$, $\alpha=1.5$, $I_{f} = 0.26$, and $\Omega_{f} = 0.4$.}
	\label{fig_1}
\end{figure}

\section{Measure of extreme events}\label{def}
A unique definition of extreme events is yet to be available, in the literature, due to its wide variety of expression in diverse areas and techniques used in research of this interdisciplinary science \cite{mcphillips2018defining, chowdhury2021review}. A general consensus has been reached by defining events as extreme events, if their occurrences are generally rare but recurring and these events exceed a predefined threshold. The measure of threshold is different as suitable to the particular discipline of science, from oceanography to climatology. In oceanography, a wave is called as rogue wave (extreme event) whenever the wave height (distance from trough to crest) exceeds eight times of the standard
deviation of the surface elevation \cite{kharif2008rogue, dysthe2008oceanic}. The $99\%$ percentile measure \cite{mcphillips2018defining}, widely used in environment science, is also used as the threshold as extreme event qualifier, but we experienced that it gives overestimates from the dynamical system perspective \cite{mishra2020routes}. The most common practice in dynamical system studies is to classify an event as extreme when it crosses a threshold height $h$=$\mu+d\sigma~(d\in \mathbb{R}\setminus\{0\}\})$, where $\mu$ is the sample mean of the data set of events (peak of a time signal), and $\sigma$ is the standard deviation \cite{chowdhury2021review, kharif2008rogue, reinoso2013extreme}.  The choice of $d$ decides the extent of the deviation from the mean and it is system dependent and arbitrarily chosen. If a higher value of $d$ is selected, $d$ determines the rarity of extreme events. This choice of $d$ is made more justified by the probability distribution of events where the threshold mark $h$ indicates which large events are really rare in occurrence. 
For our current case study, the rare large events $y_{max}$ (local maxima of ${y_i}$) are effectively characterized as extreme events when they exceed a threshold height $h=\mu+d\sigma$, where $d=6$ and, the mean and standard deviation are estimated from a sufficiently long time-series of the observable $y_i$.

\begin{figure}[h]
	\centerline{
		\includegraphics[scale=0.52]{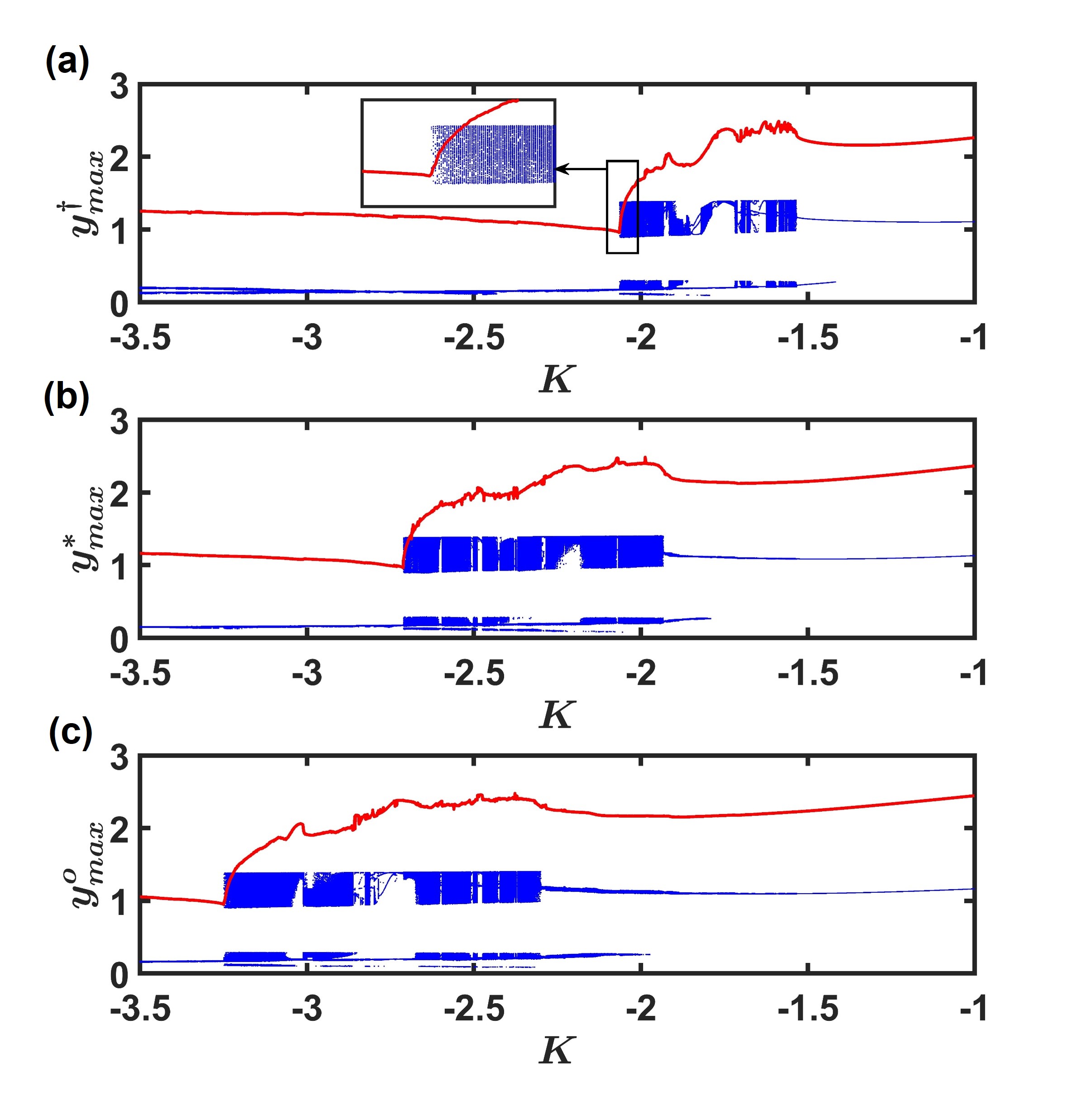}}
	\caption{Bifurcation diagrams of the observables associated with three nodes of different degrees $67$, $55$, and $47$, respectively, against repulsive coupling. We plot of local maxima of (a) $y^{\dagger}$, (b) $y^{*}$, and (c) $y^{o}$ and the threshold $h$ (red line) against $K$. Using the threshold height  $h$, extreme events are identified in three nodes, for their e respective ranges of $K$, (a) (-2.064, -2.046), (b) (-2.711, -2.692) and (c) (-3.249, -3.219). The oscillators encounter a transition from high-amplitude oscillation (rotation) to low amplitude oscillation (libration) against $|K|$. Extreme events appear near the critical value of $K$ where transition between rotation to libration occurs, in all three cases. Inset in (a) is a zoomed version near the transition from rotation to libration. For weaker coupling, each oscillator starts with periodic rotational motion and becomes aperiodic when the repulsive coupling strength increases.}
	\label{fig_2}
\end{figure}

\section{Generation of extreme events }\label{sec3}
	
\par The dynamics of all the oscillatory nodes in the network are monitored. The variable $y_i$ of the $i$-th oscillator is selected as our {\it observable}, and the local maxima of ${y_i}$ are considered as  {\it events}.
The dynamics of all oscillators exhibit the large amplitude rotation for weak interactions. But the dynamics of the largest degree node changes its nature first for a weak repulsive interaction $|K|$ and, near a critical value, switches over to small amplitude libration. Still, the dynamics shows a tendency of occasional return to large amplitude rotation due to the intrinsic instability of dynamics near the transition point. 
For increasing the strength of repulsive  interaction, one after another lower degree node is affected by  a transition to libration, but with occasional large amplitude rotation. Interestingly, the rotational motion at the previous larger degree node is then completely suppressed one after another with increasing repulsive interaction and the dynamics of the corresponding oscillators exhibit libration. These occasional transition from libration to large amplitude rotation for a short duration appears as large spikes in the temporal dynamics of $y_i$.
We inspect the possibility of the formation of these occasional large events at the oscillators by varying the repulsive coupling strength whose effects are considered as a manifestation of extreme events.
\par 
The effect of topological heterogeneity or a variation in degree of the nodes on the network dynamics  under repulsive interactions is the main focus of our study. A view of the network structure is presented in Fig.\ \ref{fig_1}(a), where the nodes are denoted by circles of varying size and with a number that depicts the degree of the node. The larger a circle, higher is the degree of the node. We first narrate our main observation on the formation of extreme events on the nodes and the role of repulsive interactions and later, describe the temporal dynamics of extreme events. Results for $5\%$ of nodes with successive higher degree \{$33, 36, 37, 39, 47, 51, 51, 55, 56, 61, 67$\} are presented in Fig.\ \ref{fig_1}(b) where the nodes are arranged in ascending order of degree along the ordinate against repulsive strength $K$. The width of the vertical color bars indicates the range of $K$ for a particular where extreme events appear; they disappear outside this range in the node.  The largest degree node ($67$) is seen most vulnerable and succumbs first for a narrow range (a thin color bar in black) of weak repulsive coupling around $K\approx -2.0$ with the formation of extreme events when all the other nodes are not affected. With an increase in repulsive interaction ($K\approx -2.25$), the next higher degree node ($61$) is only affected with the formation of  extreme events, but the highest degree node ($67$)  now stops displaying extreme events and remains silent along with the other nodes. This process of triggering the successive higher degree nodes ($56, 55, 51$, and so on) continues with increasing repulsive coupling while all the other nodes are free from extreme events. In other words, the appearance of extreme events changes position from one high degree node to the next high degree node when the strength of repulsive coupling is increased. A larger repulsive interaction strength is necessary to trigger extreme events in the low degree nodes. We mention here an anomaly that two pair of nodes with identical degrees ($51$ and $39$) show  formation of extreme events at different ranges of $K$. An initial check confirms that each pair of identical degree nodes have a variation in the sum of the degrees of the adjacent nodes. We have observed that the larger the sum of degrees of the adjacent nodes, it provokes an early appearance of extreme events at a lower range of repulsive strength. 
Furthermore, highly centralized nodes (eigenvector centrality) can be affected  (not shown here) by extreme events for lower repulsive coupling compared to lower centralized nodes though the degrees of the nodes is identical about this anomaly. 
We need further rigorous study to make any conclusive statement.

For a confirmation of the general scenario presented in Fig.\ \ref{fig_1}(b) on the formation of extreme events at the nodes of varying degree, we present three exemplary cases for three arbitrarily selected nodes of degree $67$, $55$, and $47$. We plot bifurcation diagrams of the respective observables against $K$ in Fig.\ \ref{fig_2} to identify the range of repulsive strength where extreme events appear. The local maxima of the three observables, $y^{\dagger}_{max}$, $y^{*}_{max}$, and $y^{o}_{max}$ corresponding to the nodes of degree $67$, $55$, and $47$, respectively,  are plotted for a range of $K \in [-3.5, -1]$ in Figs.\ \ref{fig_2}(a)-(c). The dynamics of the three nodes first show high amplitude periodic motion (rotation) for weak repulsive coupling $|K|$, then suddenly transits to large amplitude chaotic motion (rotation, dense blue profile) for increasing repulsive coupling and finally encounters a further transition from high amplitude chaos (rotation) to low amplitude periodic motion (libration) at a larger critical value of repulsive coupling. This transition point from chaos (rotation) to low amplitude periodic motion (libration) shifts to the left (for higher repulsive coupling) with the degree of the nodes (cf. Figs.\ \ref{fig_2}(a)-(c)). For the highest degree node, a lower critical $|K|$ is obviously a necessity for a transition to low amplitude libration, as shown in Fig.\ \ref{fig_2}(a) that corroborates the fact that the extreme events start appearing near this transition point (marked by a box near $K\approx -2$, zoomed version in the inset) for a weak repulsive interaction as suggested for the high degree node in Fig.\ \ref{fig_1}(b). The node of largest degree $67$ is affected by a sudden transition from high amplitude oscillation to low amplitude oscillation for a weaker repulsive coupling (See inset of Fig.\ \ref{fig_2}(a)).  Near this transition point at a critical value of $K$, an intrinsic instability of dynamics always occurs that leads to a tendency of occasional switching to chaotic rotation from  periodic libration before a final suppression of dynamics to libration due to increasing repulsive interactive effect. And this occasional transition to rotation appears as large spiking events and indicates the appearance of extreme events at the node. A threshold line of $h$ (red line) is drawn against $K$ in all the plots that capture a narrow range of coupling near the transition points for all the three cases, where the switching from low amplitude (libration)  to high amplitude oscillation (rotation) in a node is really occasional that appear as extreme events. We confirm this occasional transition between the two states with their temporal dynamics later. An increasingly larger critical $|K|$ is seen as needed for the transition at the lower degree nodes, as shown in Fig.\ \ref{fig_2}. 
Figure\ \ref{fig_2} clearly shows that the transition point shifts from right to left with decreasing degree of the nodes, and it confirms the general scenario that a stronger repulsive coupling $K$ is necessary for the low degree nodes to be affected by the triggering of extreme events. In a network of identical dynamical units, the underlying role of heterogeneity of degree of the nodes is thus undeniable that influences the preferential choice of a node for the formation of extreme events under repulsive interaction.

\begin{figure}[h]
	\centerline{
		\includegraphics[scale=0.095]{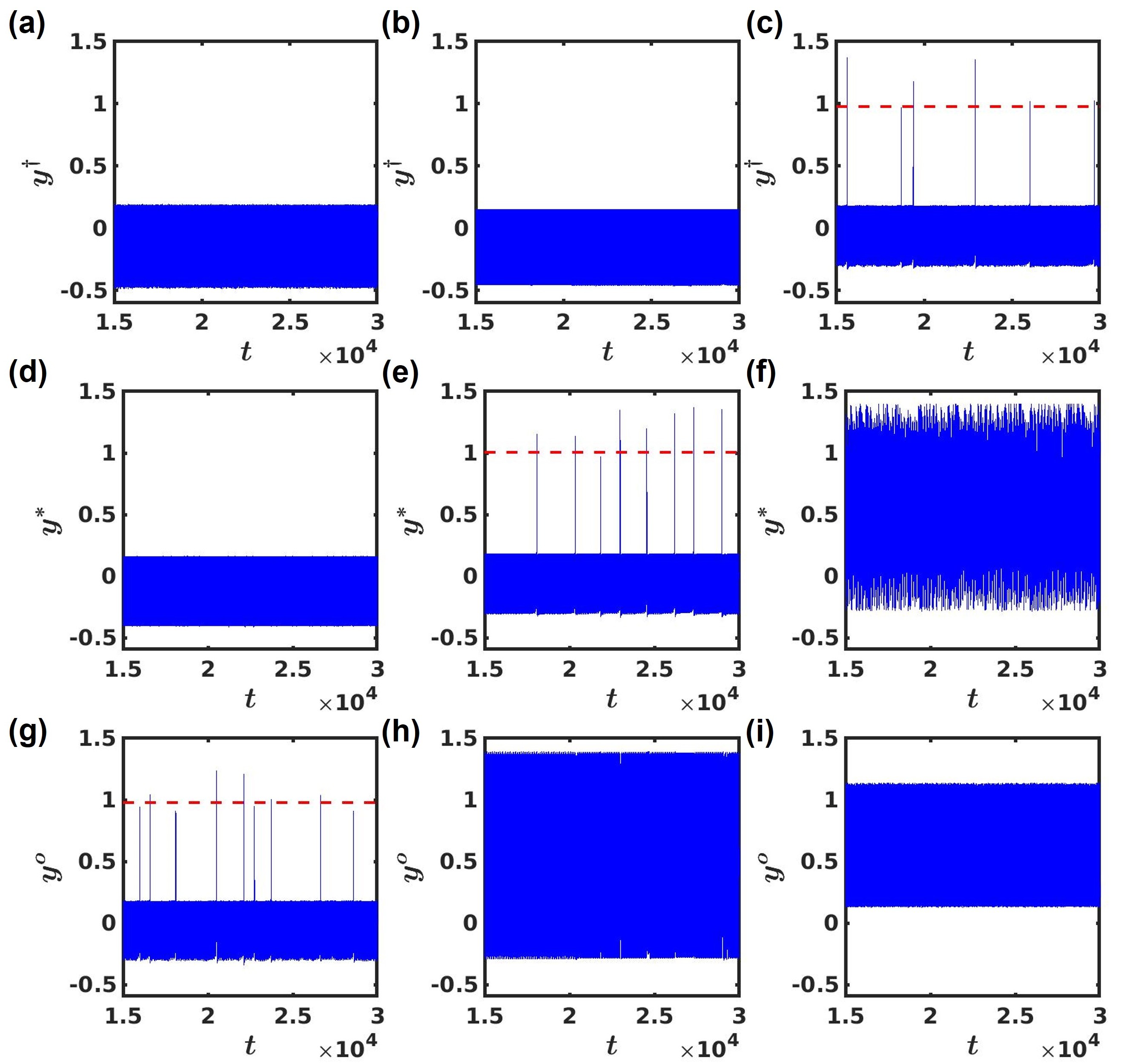}}	
	\caption{Temporal dynamics of the observables at three different nodes for different repulsive coupling strengths. Time evolution of the variables (a)-(c) $y^{\dagger}$, (d)-(f) $y^{*}$, (g)-(i) $y^{o}$ associated with three nodes with different degrees $67$, $55$, and $47$, respectively. Extreme events are displayed in (c), (e) and (g) for the respective nodes as form of occasional large spikes. Horizontal red dashed lines indicate threshold heights, as estimated separately for each case, that helps identify which large events cross the threshold and qualify as extreme events. The threshold lines are drawn only when we have observed occasional switching from libration to rotation in the time series showing signature of extreme events in the temporal dynamics of the observables. Small amplitude librational motion is observed in (a, b, d), and large amplitude rotational motion is seen in (f, h, i). There is no question of appearance of extreme events since events exist more frequently. Coupling strength: $K=-3.247$ for (a, d, g), $K=-2.7118$ for (b, e, h), and $K=-2.064$ for (c, f, i).}
	\label{fig_3}
\end{figure}

	\begin{figure*}[ht]
	\includegraphics[scale=0.132]{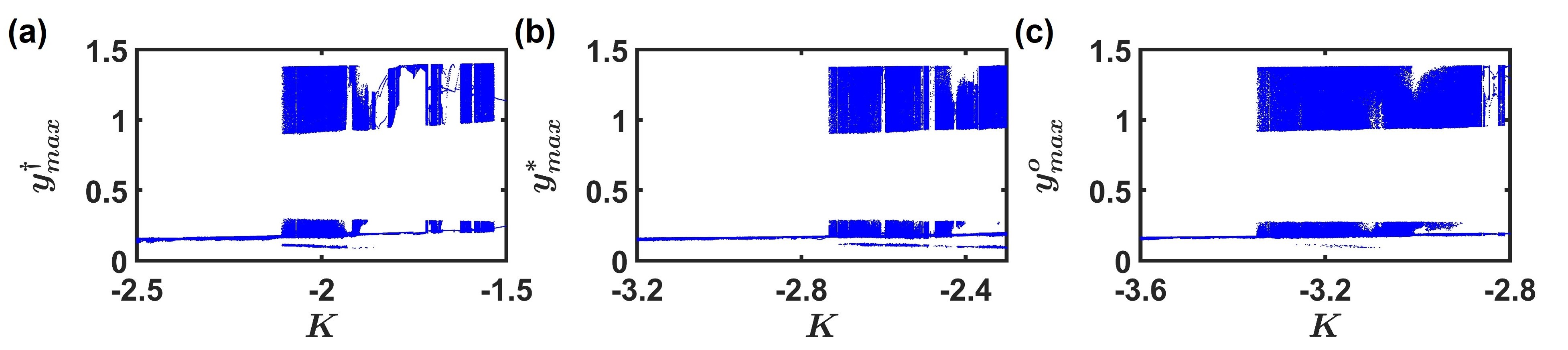}
	\caption{ Bifurcation diagrams of the dynamics of three selective nodes against repulsive coupling strength, $K$. 
		Local maxima of $y^{\dagger}$ (a),  $y^{*}$ (b), and $y^{o}$ (c) are plotted against $K$ where associated degree class ($S$) of the nodes are chosen as $67$, $55$, and $47$, respectively. The bifurcation diagrams are drawn separately using the reduced model described by Eq.\ \ref{eq.3}.
	}
	\label{fig_5}
\end{figure*}

\par The temporal patterns of evolving dynamics ($y^{\dagger}_{max}$, $y^{*}_{max}$, and $y^{o}_{max}$) in the three exemplary nodes of degree $67$, $55$, and $47$  demonstrate the dynamical character of extreme events near the transition points as depicted in the bifurcation diagrams in Figs.\ \ref{fig_2}(a)-(c). The temporal dynamics of  $y^{\dagger}$, $y^{*}$, and $y^{o}$ are presented in Fig.\ \ref{fig_3} for three selective coupling strengths, $K=-3,247$, $-2.7118$ and $-2.064$, respectively, for the three nodes. 
The highest degree node ($67$) shows extreme events for the weakest repulsive coupling $K=-2.064$ in Fig.\ \ref{fig_3}(c) with a manifestation of six intermittent large spikes in the temporal behavior of $y^{\dagger}$ when four of them (maxima of $y^{\dagger}$) exceed a threshold height $h\approx0.976$ (horizontal red dashed line). The dynamics of $y^{\dagger}$ shows libration (bounded to small amplitude oscillation), most of the time, but occasionally transits to high amplitude rotation 
resulting in intermittent large spikes and thereby leading to the formation of extreme events at the node. Figures\ \ref{fig_3}(a)-(b) do not exhibit extreme events in the dynamics of the same node (degree $67$) for stronger repulsive coupling, $K=-2.7118$ and $K=-3.247$. The amplitude of $y^{\dagger}$ remains confined to small amplitude libration, which we call as a benign or a silent state with no extreme events. 
On the other hand, the time evolution of $y^*$ at the node with a degree $55$ shows no extreme events for the weakest repulsive coupling $K=-2.064$ as displayed here in Fig.\ \ref{fig_3}(f). Instead, we observe very irregular high amplitude oscillations (rotation) that never qualify as extreme events. In a sense, the ambient amplitude is large for a long time with no sudden and occasional changes and any indicative signature of extreme events is missing here. A stronger repulsive coupling $K=-2.7118$ induces extreme events in this lower degree node ($55$), as shown in Fig.\ \ref{fig_3}(e), which appear as intermittent large spikes that indicate occasional transition from libration to rotation. Some of the intermittent large spikes of $y^*$ are larger than a threshold height $h \approx 1$ (horizontal red dashed line), confirming our claim of emergent extreme events at the node of degree $55$. For a further increase in repulsive strength, extreme events are absent, as shown in  Fig.\ \ref{fig_3}(d) for $K=-3.247$ that reveals confinement to librational motion (silent state) of the node for all the time. Extreme events are absent in both the nodes of degree $67$ and $55$ for this range of $K$.  
Finally, we focus on the third node of our choice with a degree $47$ and draw the respective time evolution of $y^{o}$ for $K=-3.247$, $-2.7118$, and $-2.064$ in Figs.\ \ref{fig_3}(g)-(i), respectively. The temporal dynamics shows formation of extreme events in Fig.\ \ref{fig_3}(g) for $K=-3.247$ in this low degree node. Most of the time, the dynamics remain in low amplitude libration, but occasionally transit to rotation when a few intermittent large events are seen crossing a threshold height ($h \approx 0.975$, horizontal red dashed line). For $K=-2.064$ and $-2.7118$, no extreme events are observed in the  node where the temporal dynamics remain confined to rotational motion all the time, as shown in Figs.\ \ref{fig_3}(h)-(i). 
It is evident from our initial inspection of Fig.\ \ref{fig_3} that an interplay of repulsive coupling strength and the degree of a node of the network decides the formation of extreme events in a particular node. A preferential choice of a node based on its degree decides the range of repulsive coupling to initiate a triggering of extreme events. High degree nodes are more vulnerable to weak repulsive coupling leading to the formation of extreme events. The low degree nodes are less susceptible to the formation of extreme events until a larger repulsive coupling strength $|K|$ is applied. We make a more convincing statement about the mechanism of the origin of extreme events using annealed network approximation in the next section.

\section{Annealed network approximation}\label{sec4}
For a better understanding of the interplay of degree heterogeneity and the repulsive mean-field interaction for the origin of extreme events, we employ the annealed network approximation  \cite{dorogovtsev2008critical, coutinho2013kuramoto} for the large network configuration model.
%
Using this approximation, the ensemble average of the network connectivity structure can be captured as $\langle A_{ij} \rangle_{ens} = \dfrac{S_i S_j}{n \langle S \rangle}$ \cite{kiss2017mathematics, kundu2017transition, subrata}, where $S_i$ is the degree of $i$-th oscillator and $\langle S \rangle=\Big(\dfrac{\sum_{i=1}^{n} S_i}{n}\Big)$ is the mean degree of the network. The coupled dynamical model in Eq.\ \ref{eq.1} is thereby reduced for $m$ number of degree classes and the equation of $l$-th class of degree $S_l$ can be written as,
\begin{equation}
\begin{array}{l}\label{eq.2}	
			{{\dot{\phi}}_{l}}  = y_l,\\
			\dot{y}_{l} = I - \sin \phi_{l} - \alpha y_{l} + I_f\sin (\Omega_f t) + \dfrac{K S_l}{n} y_{\rm{eff}},\\
			~~~~~~~~~~~~~~~~~~~~~~~~~~~~~~~~~~~~~~~~~~~~~~~~~~~~~l=1, 2, \cdots, m,
		\end{array}
	\end{equation}
	where $y_{\rm{eff}}=\dfrac{\sum_{j=1}^{n} S_j y_j}{n \langle S \rangle}$, which represents the mean-field effect that acts equally on all the oscillators of the network. More specifically, the dynamics of the oscillator corresponding to the node of degree class $S$ that represents a set of nodes is governed by,
	\begin{equation}
		\begin{array}{l}\label{eq.3}	
			\ddot{\phi}_{S} + \sin \phi_{S} + \alpha \dot{\phi}_{S} = I + I_f\sin (\Omega_f t) + \dfrac{K S}{n} y_{\rm{eff}}.
		\end{array}
	\end{equation} 
Now, we choose three exemplary nodes of degree $67, 55$, and $47$, respectively, for comparison with the original dynamics of the oscillators, described by Eq.\ \ref{eq.1}. We use the reduced model defined by Eq.\ \ref{eq.3} to draw the bifurcation diagrams in Figs.\ \ref{fig_5}(a)-(c) with a plot of local maxima $y_{max}$ against $K$ for the three selective nodes. We find a similar shift in the critical point with a change in the degree of a node as seen in Fig.\ \ref{fig_2} which is produced there using Eq.\ \ref{eq.1}. 
The bifurcation diagrams of the reduced model confirm that as we move from a high degree node to a low one, a larger repulsive strength $|K|$ is a necessity  for the transition. The shift in the critical value of transition is confirmed from the reduced model Eq.\ \ref{eq.3} for  the nodes of degree class $S$. 
\par The phase model described in Eq.\ \ref{eq.3} has a damping parameter $\alpha$ that is connected to the phase velocity $\dot \phi_S$. An additional damping term arrives at the nodes of degree class $S$ due to the influence of the repulsive mean-field interaction $K$. It appears with a multiplicative term $KS$ and connected to the effective phase velocity $y_{\rm{eff}}$. The rotational dynamics of an individual oscillator for the selected set of parameters is suppressed to libration due to an increasing  damping force.
The product $KS$ makes a combined effect for increasing the additional damping force whenever we increase either of them ($K$ or $S$), and thereby they play a crucial role in the suppression of rotation, resulting the transition from rotation to libration. Clearly when the degree class $S$ of a node is large, a smaller $K$ triggers the transition and vice versa. Near a critical value of the transition, an intrinsic instability generally occurs in dynamical systems. In particular, the dynamics undergoes a transition to libration near a critical value of the bifurcation parameter, in our case, but occasionally exhibits rotation with the formation of large spiking events (in the temporal dynamics of the observable) that are larger than threshold height. These occasional large spiking events are referred to as extreme events here. We come to a conclusion with this approximate reduced model that the transition between rotation and libration occurs due to a combined effect of the degree heterogeneity in the network topology and the repulsive coupling strength. And it is evidenced that extreme events emerge due to such occasional transitions.

\section{Statistical properties of extreme events}\label{sec5}
\par
\begin{figure}[h]
		\centerline{
			\includegraphics[scale=0.21]{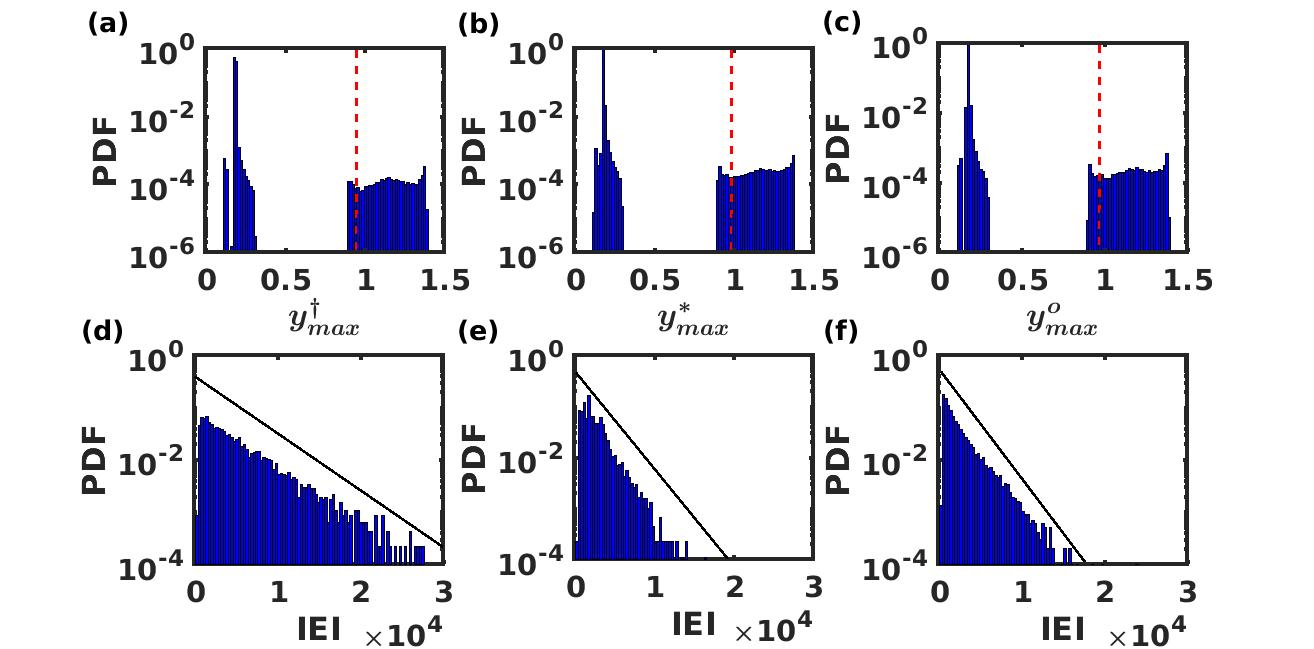}}
		\caption{Probability density function (PDF) of events and inter-event intervals (IEI) in semi-log scale. (a)-(c) PDFs of events exhibit non-Gaussian. The arrival of extreme events is indicated by the tail of the histograms in semi-log scale, specifically on the right side of the threshold. Vertical dashed red lines indicate the threshold ($h$). (d)-(f) PDFs of IEIs show a decaying nature when we plot in a semi-log scale. The black lines are proportional to the function of exponential distribution, $\exp(-\lambda~{\rm IEI})$, where, $\lambda= 0.0002263$ for (d), $\lambda=0.0004433$ for (e), and $\lambda=0.0004822$ for (f). Coupling strength: $K=-2.064$ for (a, d), $K=-2.7118$ for (b, e), and $K=-3.247$ for (c, f).}
		\label{fig_4}
	\end{figure}

  When the system dynamics frequently evolves within a range of amplitude (large or small) for a long time, the probability density of events is expected to obey a near Gaussian distribution. The appearance of extreme events in the dynamics introduces an asymmetry in the probability distribution of events, making it non-Gaussian. We plot the histograms of events for estimating the probability densities of all the events. The appearance of infrequent large events or extreme events is recognized from the tail of a distribution that indicates rarity in occurrence of extreme events. The local maxima of an observable are denoted here as events. The probability density functions (PDF) of local maxima of the observables, $y^{\dagger}_{max}$, $y^{*}_{max}$, and $y^{o}_{max}$ are plotted in Figs.\ \ref{fig_4}(a)-(c), respectively, taking a long run of the temporal dynamics as shown in Figs.\ \ref{fig_3}(c), (e), and (g). The blank windows in the PDFs indicate the absence of any intermediate amplitude events, a characteristic feature of this phase model \cite{ray2020extreme}.  Our choice of a threshold $h=\mu+6\sigma$ appears justified when we find the low probability large events exist beyond the vertical threshold lines (red dashed lines). 
	
	\par The probability distributions of arrival time interval of extreme events are shown in Figs.\ \ref{fig_4}(d)-(f) as PDF of inter-event-interval (IEI) that corresponds to the time evolution of $y^{\dagger}$, $y^{*}$, and $y^{o}$ shown in Figs.\ \ref{fig_3}(c),(e),(g), respectively. A monotonic decreasing trend in all the histograms in Figs.\ \ref{fig_4}(d)-(f). 
	 We also fit the histogram with the black line that is proportional to the PDFs of exponential distribution in Figs.\ \ref{fig_4}(d)-(f).
	The exponential distribution \cite{santhanam2008return} is described as, 
	
	\begin{equation}\label{eq.4}
		\begin{array}{lcl} F(x;\lambda)=
			\begin{cases} 
				\lambda e^{-\lambda x}; &  ~x \ge 0,  \\
				0; & ~x <0,
			\end{cases}
		\end{array}
	\end{equation}
	where $\lambda >0$ is the rate parameter.
	We calculate the coefficient of variation (CV) \cite{chowdhury2021extreme} that quantifies the ratio of standard
	deviation and mean of a data set.  Theoretically, CV$=1$ for the exponential distribution. We calculate numerically CV$=0.9063, 0.7216,$ and $0.9772$ for the data sets used in Figs.\ \ref{fig_4}(d)-(f), respectively. So, from our estimation of CV, we can conclude that PDFs of IEIs do not exactly follow exponential distribution. However, we conclude that the probability of getting long-time interval between two successive extreme events is low.

\section{Conclusion}\label{sec6}
A degree distribution as a heterogeneity in network topology is shown to induce instability in the dynamics of the oscillators leading to the formation of extreme events when the dynamical systems are influenced by repulsive mean-field interaction.  The largest degree node is most susceptible to the formation of extreme events since extreme events start triggering such oscillators for weak repulsive coupling. The low degree nodes are less vulnerable since they can only be triggered with the formation of extreme events for a stronger repulsive interaction.
We observe such behavior using a paradigmatic second-order phase model that represents the local dynamics of each node of the network.
Extreme events appear as occasional large-size spikes in the temporal dynamics of an oscillator associated with a node of the network when the dynamics occasionally switch from small amplitude to large amplitude oscillation. 
A threshold is used to detect the appearance of extreme events where the local maxima of temporal dynamics are considered as events. The distribution of all the events' height in a long temporal data record follows non-Gaussian statistics confirming the infrequent appearance of extreme events at the tail of the distribution. 
\par We are able to discern the generation of extreme events using annealed network approximation. It confirms our claim that an interplay of repulsive strength of interactions among the oscillators and 
	the degree heterogeneity leads to the origin of extreme events. A combined effect of both induces an additional damping force to each node that increases when either of them is increased. The dynamics of an oscillator associated with a high degree node thus switches from high amplitude rotation to low amplitude libration by enhanced damping for increasing repulsive interaction at a lower critical value of the coupling strength. Near this critical value, the dynamics shows instability when it switches from libration to rotation infrequently. This occasional transition from libration to rotation with a manifestation of rare large spiking events. The lower degree nodes are susceptible to such instability for a larger repulsive interaction.
\par One may wonder what will be the ideal way to control the emergence of extreme events in heterogeneous complex networks. For instance, the critical infection rate in epidemics spreading phenomena can be controlled if the highly connected nodes are cured at the early stage of spreading \cite{dezsHo2002halting}. 
In a similar fashion, extreme events in our proposed network can possibly be avoided if two or three hubs are carefully ``firewalled"  by sustaining their ``pure" rotational motion for any coupling strength. This feature may be explored in the near future.
\\

\section*{ACKNOWLEDGMENTS}
The authors would like to thank Gourab Kumar Sar for helpful discussion. TK is supported by the National Science Centre, Poland, OPUS Program Project No. 2018/29/B/ST8/00457. DG is supported by Science and Engineering Research Board (SERB), Government of India (Project no. CRG/2021/005894). 

\section*{DATA AVAILABILITY}
The data that support the findings of this study are available
within the article.

\section*{References}
\bibliographystyle{iopart-num}
\bibliography{references_ee}

\end{document}